\documentclass{jpp}

\usepackage{color}
\usepackage{graphicx}
\usepackage{natbib}

\ifCUPmtlplainloaded \else
  \checkfont{eurm10}
  \iffontfound
    \IfFileExists{upmath.sty}
      {\typeout{^^JFound AMS Euler Roman fonts on the system,
                   using the 'upmath' package.^^J}%
       \usepackage{upmath}}
      {\typeout{^^JFound AMS Euler Roman fonts on the system, but you
                   dont seem to have the}%
       \typeout{'upmath' package installed. JPP.cls can take advantage
                 of these fonts, if you use 'upmath' package.^^J}%
      }
  \else
  \fi
\fi

\ifCUPmtlplainloaded \else
  \checkfont{msam10}
  \iffontfound
    \IfFileExists{amssymb.sty}
      {\typeout{^^JFound AMS Symbol fonts on the system, using the
                'amssymb' package.^^J}%
       \usepackage{amssymb}%

      }{}
  \fi
\fi

\ifCUPmtlplainloaded \else
  \IfFileExists{amsbsy.sty}
    {\typeout{^^JFound the 'amsbsy' package on the system, using it.^^J}%
     \usepackage{amsbsy}}
    {\providecommand\boldsymbol[1]{\mbox{\boldmath $##1$}}}
\fi

\newsavebox{\astrutbox}
\sbox{\astrutbox}{\rule[-5pt]{0pt}{20pt}}

\title[Building a weak shockwave from linear modes]{Building a weak shockwave from linear modes}

\author[A. Bret and R. Narayan]%
{Antoine Bret$^{1,2}$, Ramesh Narayan$^3$%
  \thanks{Email address for correspondence: antoineclaude.bret@uclm.es}
}

\affiliation{$^1$ETSI Industriales, Universidad de Castilla-La Mancha, 13071 Ciudad Real, Spain\\[\affilskip]
$^2$Instituto de Investigaciones Energ\'{e}ticas y Aplicaciones Industriales, Campus Universitario de Ciudad Real, 13071 Ciudad Real, Spain\\[\affilskip]
$^3$Harvard-Smithsonian Center for Astrophysics, Harvard University, 60 Garden St., Cambridge, MA 02138 USA
}

\date{?; revised ?; accepted ?. - To be entered by editorial office}

\begin{document}

\maketitle

\begin{abstract}
In shockwave theory, the density, velocity and pressure jumps are derived from the conservation equations. Here, we address the physics of a weak shock the other way around. We first show that the density profile of a weak shockwave in a fluid can be expressed as a sum of linear acoustic modes.  The shock so built propagates at the speed of sound and matter is exactly conserved at the front crossing. Yet, momentum and energy are only conserved up to order 0 in powers of the shock amplitude. The density, velocity and pressure jumps are similar to those of a fluid shock, and an equivalent Mach number can be defined. A similar process is possible in magnetohydrodynamic. Yet, such a decomposition is found impossible for collisionless shocks due to the dispersive nature of ion acoustic waves. Weakly nonlinear corrections to their frequency do not solve the problem. Weak collisionless shocks could be inherently nonlinear, non-amenable to any linear superposition. Or they could be nonexistent, as hinted by recent works.
\end{abstract}

\maketitle

\section{Introduction}
In a neutral fluid or in magnetohydrodynamic (MHD), the density, velocity or pressure jumps of a shockwave can be straightforwardly determined from the  conservation equations \citep{Zeldovich,Kulsrud2005}. In contrast, theoretical works dealing with shocks density profile or corrugations, are analytically more involved because they need to go beyond the conservation laws and deal with the proper dynamics of the fluid \citep{LandauFluid,WouchukPRE2001,LemoineApJ2016}.

The goal of this article is to contribute to the second kind of studies by proposing an analytical theory of weak shocks which could prove useful, for example, when studying the interaction of a shock with an obstacle \citep{Chaudhuri2013}. In the theory of shockwaves, the density, velocity and pressure jumps are derived from the conservation equations, while the shock profile is secondary. Here, we work the other way around. We start from an analytical expression of the density profile, then check the fulfilment of the conservations laws and finally derive the jumps.

The theory exposed in the following is limited to weak shocks because it relies on the following premise. In a homogenous fluid of density $\rho_0$, deviations from equilibrium of amplitude $\delta\rho \ll\rho_0$ can be analyzed using linear theory. It should then be possible to consider a weak shockwave of amplitude $\delta\rho$ as a weak perturbation of the medium, hence analyzable in terms of linear modes. In Section \ref{sec:fluid} of this article, we show that up to a good approximation, this program can be achieved in neutral fluid. In Section \ref{sec:coll-less}, we find it is impossible to achieve in a collisionless ion/electron plasma. In the conclusion, we stress that what is achieved here for neutral fluids can also be achieved for MHD fluids. We also discuss the possible reasons why collisionless shocks escape the procedure.

\section{Weak shocks in fluids as a sum of linear modes}\label{sec:fluid}
Consider then a fluid at rest with homogenous density $\rho_0$. We study deviations from equilibrium of the form,
\begin{equation}
\rho(x,t) = \rho_0 + \rho_1(x,t),
\end{equation}
with $\rho_1 \ll \rho_0$. Assume now that at time $t=0$ the density profile of a weak shock has,
\begin{equation}\label{eq:profile}
  \rho_1(x,0) = \delta \rho ~ \mathrm{sign} (x),
\end{equation}
with $\delta \rho \ll \rho_0$. Can it be expressed as a superposition of linear acoustic modes?

The linear acoustic modes of a fluid are of the form,
\begin{equation}\label{eq:lmodes}
  \rho_1(x,t) = \varphi(k) \exp [ ik(x \pm c_s t)],
\end{equation}
where $c_s$ is the speed of sound in the medium. At $t=0$, Eq. (\ref{eq:lmodes}) reads $\rho(x,0) = \varphi(k) e^{ik x}$ so that at this time, the possibility of expressing the profile (\ref{eq:profile}) as a sum of linear modes simply stems for the possibility of Fourier expanding it. Fourier expansion at time $t=0$ provides $\varphi(k)=-i/k\pi$. We can therefore write,
\begin{equation}\label{eq:profilek}
  \rho(x,0) = \rho_0 + \delta \rho \int_{-\infty}^{+\infty}  \frac{-i}{k\pi} e^{ikx} dk.
\end{equation}

\begin{figure}
\begin{center}
 \includegraphics[width=.9\textwidth]{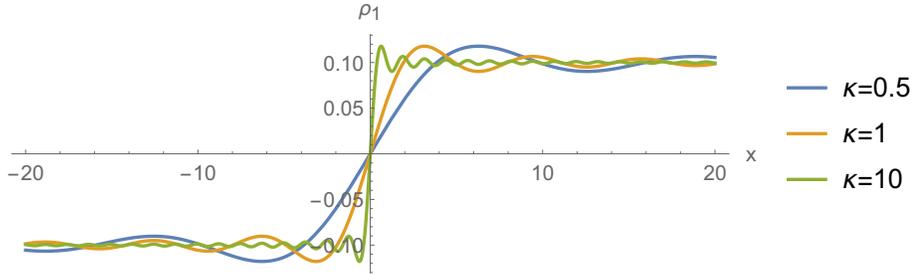}
\end{center}
\caption{Reconstruction of the shock profile from Eq. (\ref{eq:profilekappa}) with $\delta\rho=0.1$.}\label{fig:fluid}
\end{figure}

It is interesting to observe how Eq. (\ref{eq:profilek}) reconstructs the shock density profile, as an increasing number of $k$'s are accounted for. To this extent, we define,
\begin{equation}\label{eq:profilekappa}
\delta\rho^\kappa(x,0) =  \delta\rho \int_{-\kappa}^{+\kappa} \varphi(k) e^{ikx} dk.
\end{equation}
Figure \ref{fig:fluid} plots this function for 3 values of $\kappa$. It takes large $k$'s to reproduce the sharp front.

\subsection{Propagating the wave packet}
It is straightforward to propagate the profile defined by Eq. (\ref{eq:profilek}), replacing $x \rightarrow x + c_s t$ to obtain a leftward propagation. The result reads,
\begin{equation}\label{eq:profilekt}
  \rho(x,t) = \rho_0 + \delta \rho \int_{-\infty}^{+\infty}  \frac{-i}{k\pi} e^{ik(x+c_st)} dk,
\end{equation}
which is a superposition of normal modes of the form (\ref{eq:lmodes}). In addition,
\begin{equation}\label{eq:profilerho}
   \rho_0 + \delta \rho \int_{-\infty}^{+\infty}  \frac{-i}{k\pi} e^{ik(x+c_st)} dk = \rho_0 + \delta \rho ~\mathrm{sign} (x+c_st).
\end{equation}
We thus find that choosing the same $\varphi(k)$, the superposition of linear modes (\ref{eq:lmodes}) produces a propagating weak shock of amplitude $\delta \rho$.

\subsection{Velocity field}
To further study our weak shock we need to determine the velocity field. To this extent, consider the conservation of matter equation,
$$\frac{\partial (\rho_0+\rho_1)}{\partial t} + \frac{\partial (\rho_0+\rho_1) v}{\partial x} = 0,$$
we derive at first order\footnote{$v$ is first order since we work in the frame where the fluid is at rest.},
$$ \frac{\partial \rho_1}{\partial t} + \rho_0\frac{\partial v_1}{\partial x} = 0 ~~\Rightarrow ~~\frac{\partial v_1}{\partial x} = -\frac{1}{\rho_0}\frac{\partial \rho_1}{\partial t}.$$
From (\ref{eq:profilekt}) we obtain the time derivative of $\rho_1$ as,
\begin{equation}
 \frac{\partial \rho_1}{\partial t} = \delta \rho \int_{-\infty}^{+\infty}  \frac{-i}{k\pi} ikc_s e^{ik(x+c_st)} dk,
\end{equation}
so that,
\begin{equation}
\frac{\partial v_1}{\partial x} =-\frac{1}{\rho_0} \delta \rho \int_{-\infty}^{+\infty}  \frac{-i}{k\pi} ikc_s e^{ik(x+c_st)} dk,
\end{equation}
and,
\begin{eqnarray}\label{eq:profilev}
v_1 &=&  -c_s\frac{ \delta \rho}{\rho_0} \int_{-\infty}^{+\infty}  \frac{-i}{k\pi}  e^{ik(x+c_st)} dk + const,  \nonumber \\
    &=&  -c_s\frac{ \delta \rho}{\rho_0} \mathrm{sign} (x+c_st) + const.
\end{eqnarray}

Therefore, a velocity jump of amplitude $2c_s\delta \rho/\rho_0$ propagates at $c_s$, together with the density jump (\ref{eq:profilekt}).
The $const$ in this expression reflects the fact that it holds in any frame of reference.
In the following we shall work in the frame of the shock front, setting $const=c_s$.

\subsection{Conserved quantities}
Shockwave theory usually begins with the conservation of matter, momentum and energy. Here, we built the shock from normal modes, without alluding to these laws. We now check them.

To this extent, we work in the shock frame. In this frame, we have from Eqs. (\ref{eq:profilerho},\ref{eq:profilev}),
  \begin{equation}
   \rho  =  \left\{\begin{array}{r}
 \rho_0-\delta\rho,~~x<0, \\
 \rho_0+\delta\rho,~~x>0,
               \end{array} \right.
               ~~~~~~
   v  = c_s \times \left\{\begin{array}{r}
1+\delta\rho/\rho_0 ,~~x<0, \\
1-\delta\rho/\rho_0,~~x>0.
               \end{array} \right.
\end{equation}
For the conservation of matter, momentum and energy, we therefore check the conservation of the quantities $\rho v$, $\rho v^2 + P$ and $\rho v^2/2 + \gamma P/(\gamma -1)$ respectively ($\gamma$ is the adiabatic index of the fluid). That is,
\begin{enumerate}
  \item Conservation of matter
  \begin{equation}
   \rho v =  \left\{\begin{array}{r}
 (\rho_0-\delta\rho)c_s\left(1+\frac{\delta\rho}{\rho_0}\right)
                           =\rho_0c_s\left(1-\frac{\delta\rho^2}{\rho_0^2}  \right),~~x<0, \\
 (\rho_0+\delta\rho)c_s\left(1-\frac{\delta\rho}{\rho_0}\right)
                           =\rho_0c_s\left(1-\frac{\delta\rho^2}{\rho_0^2}  \right),~~x>0.
               \end{array} \right.
\end{equation}
Matter is therefore exactly conserved through the front. With $P=\rho c_s^2/\gamma$ (see eg. \cite{LandauFluid}, \S64) we check the conservation of momentum and energy.
  \item Conservation of momentum,
\begin{equation}
   \rho v^2 + P =  \left\{\begin{array}{r}
(\rho_0 - \delta\rho)c_s^2\left(1+\frac{\delta\rho}{\rho_0} \right)^2 + (\rho_0 - \delta\rho)\frac{c_s^2}{\gamma}
=\rho_0c_s^2\left( \frac{\gamma+1}{\gamma} + \frac{\gamma-1}{\gamma}\frac{\delta\rho}{\rho_0}-\left(\frac{\delta\rho}{\rho_0}\right)^2-\left(\frac{\delta\rho}{\rho_0}\right)^3   \right),~~x<0, \\
(\rho_0 + \delta\rho)c_s^2\left(1-\frac{\delta\rho}{\rho_0} \right)^2 + (\rho_0 + \delta\rho)\frac{c_s^2}{\gamma}
=\rho_0c_s^2\left( \frac{\gamma+1}{\gamma} - \frac{\gamma-1}{\gamma}\frac{\delta\rho}{\rho_0}-\left(\frac{\delta\rho}{\rho_0}\right)^2+\left(\frac{\delta\rho}{\rho_0}\right)^3\right),~~x>0.
               \end{array} \right.
\end{equation}
Momentum is therefore conserved up to order 0 in $\delta\rho$, save for the specific case $\gamma=1$ where it is conserved up to second order in $\delta \rho$. This would correspond to an isothermal gas which radiates the shock energy and maintains a constant temperature. In all other cases, momentum conservation is maintained only up to order 0 in $\delta \rho$.
  \item Conservation of energy,
\begin{equation}
   \frac{\rho v^2}{2} + \frac{\gamma}{\gamma -1} P =  \left\{\begin{array}{r}
\frac{1}{2}\rho_0 c_s^2\left( \frac{\gamma+1}{\gamma-1} + \frac{\gamma-3}{\gamma-1}\frac{\delta\rho}{\rho_0} -\left(\frac{\delta\rho}{\rho_0}\right)^2-\left(\frac{\delta\rho}{\rho_0}\right)^3 \right),~~x<0, \\
\frac{1}{2}\rho_0 c_s^2\left(  \frac{\gamma+1}{\gamma-1} - \frac{\gamma-3}{\gamma-1}\frac{\delta\rho}{\rho_0} -\left(\frac{\delta\rho}{\rho_0}\right)^2+\left(\frac{\delta\rho}{\rho_0}\right)^3 \right),~~x>0.
               \end{array} \right.
\end{equation}
Energy is therefore conserved up to order 0 in $\delta\rho$ for any $\gamma$.
\end{enumerate}

In summary,  matter is exactly conserved while momentum and energy are conserved to order 0 in $\delta\rho$. Such a discrepancy is due to the entropy jump, which vanishes in our case since it arises from the crossing of a structure eventually made up of linear Vlasov modes. However, since for a real weak shock, the entropy difference scales like $\delta\rho^3$ (\cite{LandauFluid}, \S 86), considering it is 0 does yields incorrect results for the jump of the other quantities.

\subsection{Jumps and equivalent Mach number}\label{sec:jump}
 We can write the density jump of the shock as,
\begin{equation}\label{eq:rfluid}
r = \frac{\rho_0 + \delta\rho}{\rho_0 - \delta\rho}= 1 + 2\frac{\delta\rho}{\rho_0}+\mathcal{O}(\delta\rho)^2.
\end{equation}
Conversely, the velocity jump reads,
\begin{equation}\label{eq:rfluidv}
r_v = \frac{c_s-c_s\frac{ \delta \rho}{\rho_0}}{c_s+c_s\frac{ \delta \rho}{\rho_0}} = \frac{\rho_0 -\delta \rho}{\rho_0 v_0+ \delta \rho} = \frac{1}{r}.
\end{equation}
Finally, the pressure jump reads,
\begin{equation}\label{eq:rfluidP}
r_P=\frac{(\rho_0 + \delta\rho)c_s^2}{(\rho_0 - \delta\rho)c_s^2} =r.
\end{equation}
Now, according to the shockwave theory (\cite{Thorne2017}, p. 905),
\begin{eqnarray}\label{eq:rfluidold}
r=\frac{(\gamma+1) \mathcal{M}^2}{(\gamma-1) \mathcal{M}^2+2}
                    &=& 1 + \frac{4 (\mathcal{M}-1)}{\gamma+1}+\mathcal{O}(\mathcal{M}-1)^2,  \nonumber  \\
r_P=\frac{2 \gamma^2\mathcal{M}^2}{\gamma+1}-\frac{\gamma-1}{\gamma+1}
 &=& 1 + \frac{4 (\mathcal{M}-1)}{\gamma+1}+\mathcal{O}(\mathcal{M}-1)^2,
\end{eqnarray}
where $\mathcal{M}$ is the shock Mach number. It appears that at first order in $(\mathcal{M}-1)$, the pressure and density jumps are equal. Noticing from Eqs. (\ref{eq:rfluid},\ref{eq:rfluidP}) that our built-up shock also has the same density and pressure jumps, we can assign it an equivalent Mach number $\mathcal{M}_{eq}$ by equating the Taylor expansions (\ref{eq:rfluid}) and (\ref{eq:rfluidold}),
\begin{equation}
\mathrm{define}~\mathcal{M}_{eq}/. ~~ 2\frac{\delta\rho}{\rho_0} = \frac{4 (\mathcal{M}_{eq}-1)}{\gamma+1},
\end{equation}
giving,
\begin{equation}\label{eq:Macheq}
\mathcal{M}_{eq} = 1 +  \frac{1+\gamma}{2 \gamma} \frac{\delta\rho}{\rho_0}.
\end{equation}

The shock that we just built from the superposition of linear modes mimics a weak fluid shocks with a Mach number given by Eq. (\ref{eq:Macheq}).

\section{Collisionless shocks}\label{sec:coll-less}
In a collisionless electron/proton plasma, ion acoustic waves (IAW) follow the dispersion,
\begin{equation}\label{eq:IAW}
  \omega_k = \pm k\sqrt{\frac{k_BT_e/m_p}{1+k^2\lambda_D^2}},
\end{equation}
where $T_e$ is the electronic temperature and $\lambda_D$ the Debye length. These waves are Landau damped for $k>k_m= \lambda_D^{-1}\sqrt{T_e/T_p}$, $T_p$ being the protonic temperature (\cite{Thorne2017}, p. 1047). If the shock profile is to be given by an expression of the form,
\begin{equation}\label{eq:wavepacket}
  \rho_1(x,t)=\int \varphi (k) e^{ikx-i\omega_kt} dk,
\end{equation}
then again, setting $t=0$ implies $\varphi (k)$ is the Fourier transform of  $\rho_1(x,0)$. Assuming for example a step function at $t=0$ like in Section \ref{sec:fluid}, Eq. (\ref{eq:wavepacket})  then propagates it for $t \neq 0$. After some algebra we get,
\begin{equation}\label{eq:rho1_ep}
\rho_1(x,t) =  \frac{\delta \rho}{\pi} \int_{-k_m}^{k_m}   \sin \left(kx + k t\sqrt{\frac{k_BT_e/m_p}{1+k^2\lambda_D^2}} \right)\frac{dk}{k} ,
\end{equation}
where the integration is over $[-k_m,k_m]$ instead of $[-\infty,\infty]$ to account for the Landau damping. Setting,
\begin{equation}
u = k\lambda_D, ~~~ X = x/\lambda_D, ~~~ \tau =  t\frac{\sqrt{k_BT_e/m_p}}{\lambda_D},
\end{equation}
gives
\begin{equation}
\rho_1(X,\tau) =  \frac{\delta \rho}{\pi} \int_{-\sqrt{T_e/T_p}}^{\sqrt{T_e/T_p}}   \sin \left[u \left( X + \frac{\tau}{\sqrt{1+u^2}}  \right) \right]\frac{du}{u} .
\end{equation}
Setting $X_s=X+\tau$ to follow the propagation, we get
\begin{equation}\label{eq:iawstep}
\rho_1(X_s,\tau) =  2\frac{\delta \rho}{\pi} \int_0^{\sqrt{T_e/T_p}}   \sin \left[u X_s \left( 1 + \frac{\tau f(u)}{X_s}  \right) \right]\frac{du}{u} .
\end{equation}
with,
\begin{equation}\label{eq:f}
  f(u)=\frac{1-\sqrt{1+u^2}}{\sqrt{1+u^2}}.
\end{equation}

\begin{figure}
\begin{center}
 \includegraphics[width=.5\textwidth]{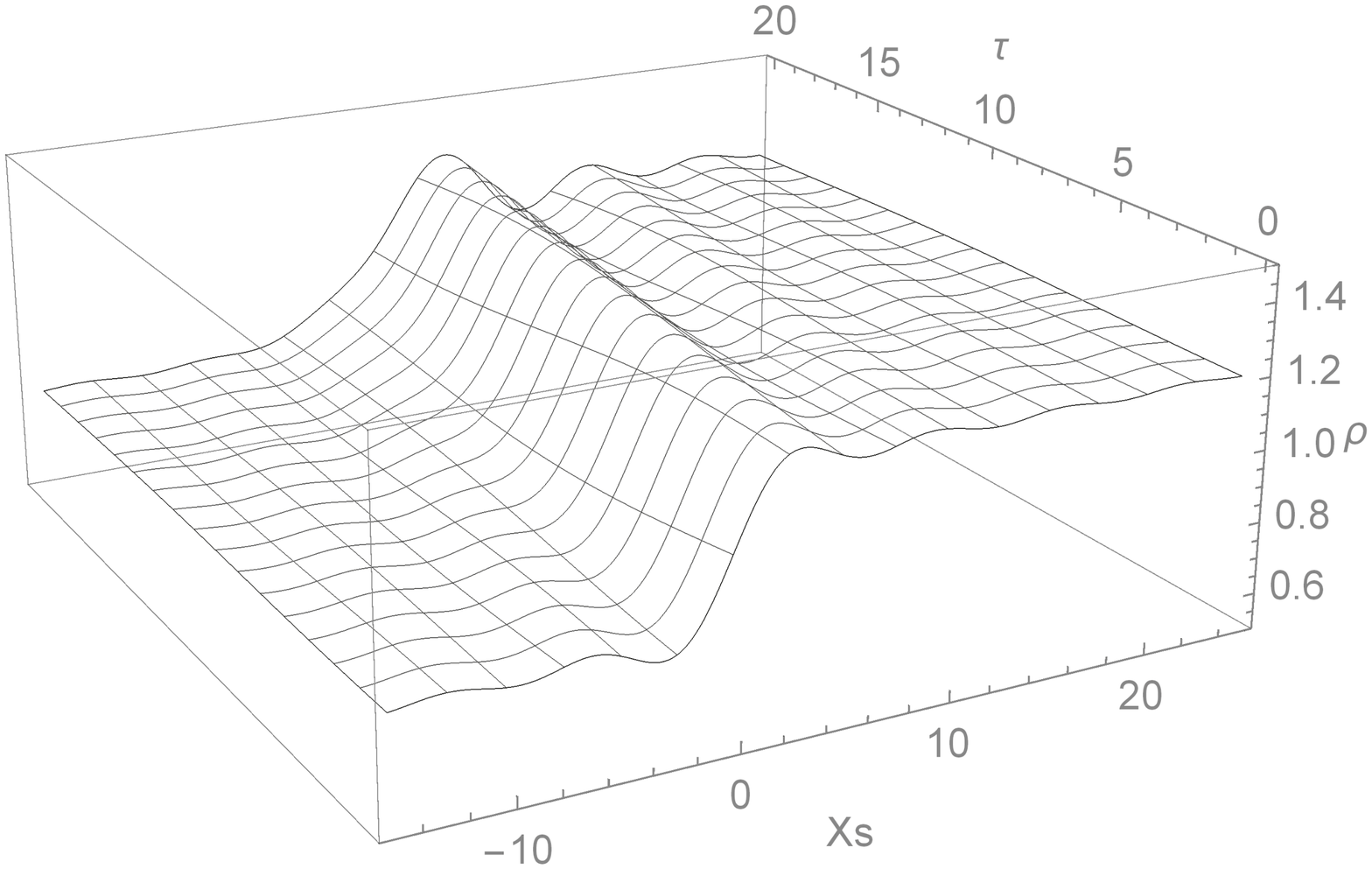} \includegraphics[width=.48\textwidth]{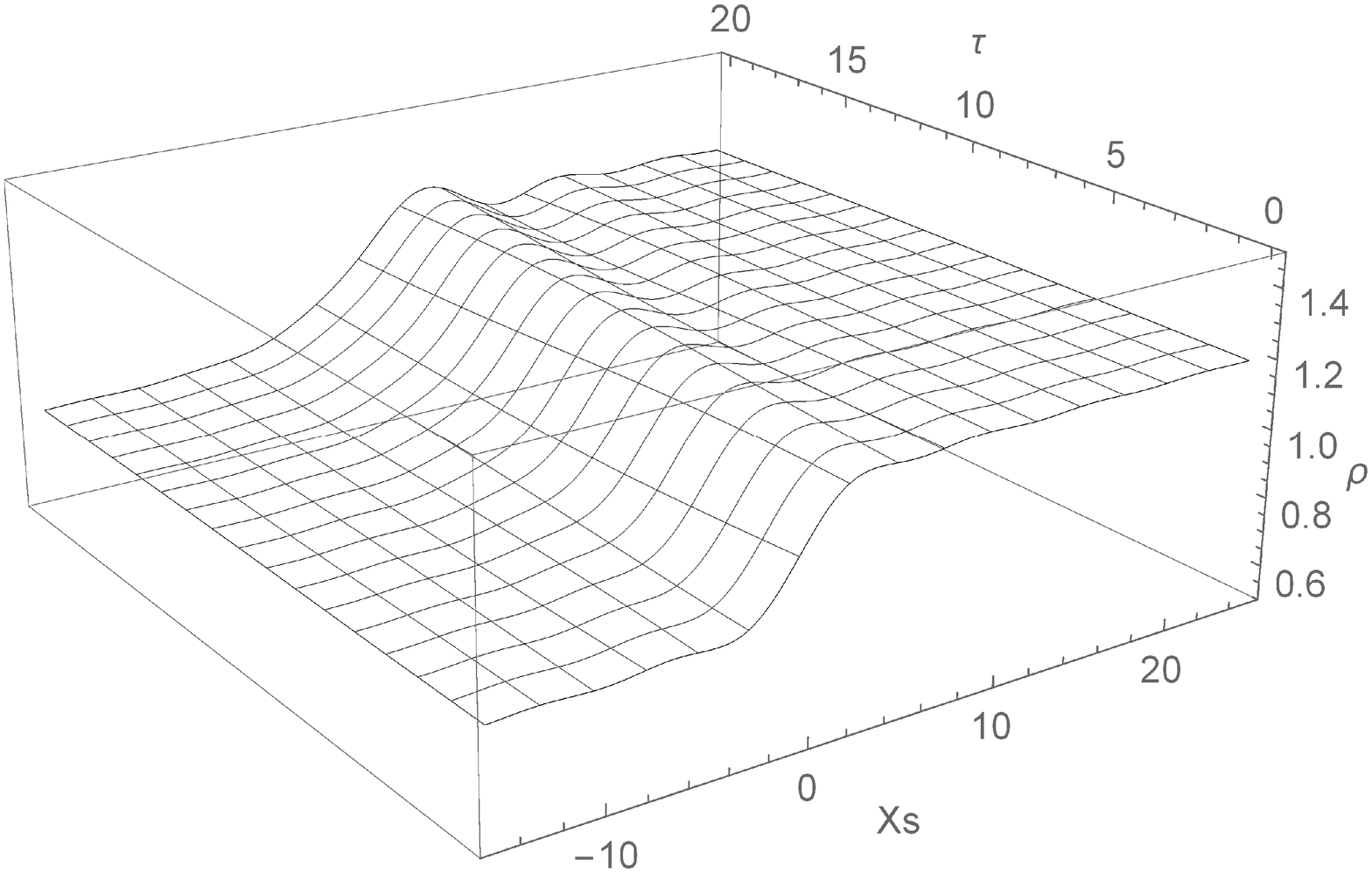}
\end{center}
\caption{Density profile $\rho(X_s,\tau) = 1+0.2\rho_1(X_s,\tau)$, for $T_e/T_p=1$. $\delta\rho$ is set to 0.2 for more clarity. Left: initial step function at $t=0$. Right: initial $\arctan (x/\lambda_D)$ function at $t=0$.}\label{fig:iaw}
\end{figure}

Equation (\ref{eq:iawstep}) is plotted on Figure \ref{fig:iaw}-left\footnote{The initial shape is not a step function because integration in Eq. (\ref{eq:iawstep}) is not performed up to $+\infty$.}. It appears that the initial shape is not propagated without distortion. Figure \ref{fig:iaw}-right shows the same calculation, but for an initial profile given by $\arctan (x/\lambda_D)$. Again, propagation distorts the profile.

Considering the form of Eq. (\ref{eq:IAW}), the patterns match qualitatively the analysis of \cite{Krall1997} (smaller wavelengths travel slower, hence trail behind the front). But  stationarity is not achieved in any frame.

\bigskip

Indeed we now prove that because of the form of the dispersion (\ref{eq:IAW}), it is impossible to build a IAW wave packet that does not distort when propagating.

Assume then a IAW wave packet of the form,
\begin{equation}\label{eq:wavepacket1}
  \rho(x,t)=\int \varphi (k) e^{ikx-i\omega_kt} dk.
\end{equation}
To see if it can result in a stationary structure in a frame moving at speed $V$, we first operate $x \rightarrow X + Vt$ with $V\in \mathbb{R}$, and form the time derivative of the resulting expression,
\begin{equation}
  \frac{\partial \rho(X,t)}{\partial t}=\int   i(kV-\omega_k) \varphi (k)   e^{ikX + i(kV-\omega_k)t}dk.
\end{equation}
If the packet (\ref{eq:wavepacket1}) is to be stationary in some frame at velocity $V$ then the expression above must vanish $\forall (X,t)$. In particular, setting $t=0$ must give,
\begin{equation}
 \int  (kV-\omega_k) \varphi (k)   e^{ikX}dk = 0,
\end{equation}
so that the Fourier transform of the function $F(u)=(uV-\omega_u) \varphi (u)$ must be 0. This implies $F(u)=0$, that is, $\omega_u=uV$ or trivially $\varphi (u)=0$.

Therefore, unless $\exists V \in R/\omega_u=uV$, it is impossible to build a IAW packet which is stationary is some frame. Conversely, if $\exists V\in \mathbb{R}/\omega_u= uV$, then any superposition of linear acoustic modes propagates without distortion since all the harmonics propagate at the same speed $V$.

Note that the problem does not come from the limited integration domain in (\ref{eq:rho1_ep}) imposed by Landau damping, but from the dispersive form of $\omega_k$ in (\ref{eq:IAW}).

\section{Conclusion}\label{sec:conclu}
We considered building up a weak shock from a superposition of linear modes. The attempt is successful for any medium where the dispersion relation is of the form $\omega_k = kV$. Conservation laws are met exactly for matter, and up to order 0, in powers of the shock amplitude, for the momentum and the energy.

As long as the linear modes of the fluid are of the form (\ref{eq:lmodes}), that is, enjoy an acoustic dispersion like $\omega(k) = k c, ~c\in \mathbb{R}$, the reasoning above holds. In particular, the dispersion of the fast, intermediate and slow MHD modes read $\omega=k c_+$, $\omega=k c_A$ and $\omega=k c_-$, where $c_A$ is the Alfv\'{e}n speed and $c_\pm$ the velocity of the fast and slow modes (\cite{Kulsrud2005}, ch. 5). The reasoning above applies therefore to the 3 kinds of weak MHD shocks, substituting $c_s$ by the corresponding velocity.

For collisionless plasma, the program cannot be realized, precisely because the dispersion relation is not of the suitable form. Non-linear frequency shifts to Eq. (\ref{eq:IAW}) have been studied by \cite{BergerPoP2013,AffolterPoP2019}\footnote{See for example Eq. (22) of \cite{BergerPoP2013}.}. Yet, they still not result in a dispersionless expression for $\omega_k$, which is a necessary condition to form a stationary profile in a moving frame, as proved in Section \ref{sec:coll-less}.

The failure of our program for collisionless shocks can have two origins:

\begin{itemize}
  \item Weak collisionless shocks are intrinsically non-linear phenomena which cannot be captured by any superposition of linear modes. This would be in line with, for example, the theory of shocks in viscous fluids, where an analytical profile can be found (\cite{LandauFluid}, \S 93) while the linear modes are all damped by viscosity (\cite{Zeldovich}, ch. 1, \S 22). Similar considerations apply to solitons in a plasma \citep{Tran1979PhyS}.

      Closely related is the long-known fact that collisionless shocks are possible because nonlinear wave steepening balances the spreading of the wave packet due to dispersion (see \cite{Tidman1971} ch. 6, \cite{Galeev1976}, \cite{Kennel1985}, \cite{Balogh2013} \S 2.1.4). Yet, introducing such effects, be it phenomenologically, would go against the main idea of our model which is precisely to reconstruct a shock from a superposition of linear modes.
  \item Extremely weak collisionless shock, with Mach number $1+\varepsilon$, cannot exist. That would be coherent with the model of \cite{BretJPP2018} for collisionless shocks in pair plasmas, where no solutions could be found for Mach numbers lower than 1.34. Note that this latter model has been found correct for strong shocks \citep{Haggerty2021}.
\end{itemize}

To our knowledge, no particle-in-cell simulations of collisionless shocks for Mach number $\mathcal{M} < 2$ have been achieved so far\footnote{See \cite{2018ApJ...858...95G, 2018ApJ...864..105H, 2019ApJ...876...79K} for $\mathcal{M} = 2$.}. Future numerical works with $1<\mathcal{M} < 2$ could help resolving the present issue.

\section{Acknowledgments}
A.B. acknowledges support by grants ENE2016-75703-R from the Spanish Ministerio
de Econom\'{\i}a y Competitividad and SBPLY/17/180501/000264 from the Junta de
Comunidades de Castilla-La Mancha. R.N. acknowledges support from the NSF Grant No. AST-
1816420. R.N. thanks the Black Hole Initiative at Harvard University for support. The BHI is funded by grants from the John Templeton Foundation and the Gordon and Betty Moore Foundation. Thanks are due to Ian Hutchinson for enriching discussions.


\end{document}